\documentstyle[12pt,psfig]{article}
\newcommand{\tb}{\mbox{{$\,\tan\beta$}}}
\newcommand{\cb}{\mbox{{$\,\cos\beta$}}}
\newcommand{\sbt}{\mbox{{$\,\sin\beta$}}}

\newcommand{\cbsq}{\mbox{{$\,\cos^2\beta$}}}
\newcommand{\sbsq}{\mbox{{$\,\sin^2\beta$}}}
\newcommand{\tbsq}{\mbox{{$\,\tan^2\beta$}}}

\newcommand{\gsim}{\mbox{ \raisebox{-1.0ex}{$\stackrel{\textstyle >}
{\textstyle \sim}$ }}}
\newcommand{\lsim}{\mbox{ \raisebox{-1.0ex}{$\stackrel{\textstyle <}
{\textstyle \sim}$ }}}

\begin{document}
\vspace*{-10mm}
\baselineskip12pt
\begin{flushright}
\begin{tabular}{l}
{\bf KEK-TH-458 }\\
{\bf KEK Preprint 95-173}\\
{\bf OU-HET 229 }\\
December 1995\\
\end{tabular}
\end{flushright}
\baselineskip24pt

\vspace{8mm}
\begin{center}
{\Large \bf Probing the heavy Higgs mass by the
measurement of the Higgs decay branching ratios in
the minimal supersymmetric standard model}\\
\vglue 8mm
{\bf Jun-ichi Kamoshita$^a$,
     Yasuhiro Okada$^a$
     and Minoru Tanaka$^b$}\\

{\it Theory Group, KEK, Tsukuba, Ibaraki 305, Japan$^a$}\\
{and} \\
{\it Department of Physics, Osaka University,
 Toyonaka, Osaka 560, Japan$^b$}\\
\end{center}
\vglue 20mm
\begin{center}
{\bf ABSTRACT} \\
\baselineskip24pt
\vglue 10mm
\begin{minipage}{14cm}
       We examine whether the parameters in the Higgs sector
       of the minimal supersymmetric standard model
       can be determined by detailed study of production cross section
       and decay branching ratios of the Higgs particle.
       Assuming that
       the lightest CP-even Higgs boson ($h$) is observed at
       a future $e^+ e^-$ linear collider with
       $\sqrt{s}=300\sim500$GeV,
       we show that the value of CP-odd scalar mass is
       determined from the ratio of the two branching ratios,
       $Br(h\to b\bar{b})$ and $Br(h\to c\bar{c})+Br(h\to gg)$,
       almost independently of the stop mass scale.
\end{minipage}
\end{center}
\vfill
\newpage
\baselineskip24pt

 In the search for the theory beyond the standard model (SM),
 the supersymmetric (SUSY) extension is considered to be an
 attractive and promising candidate.
 It is, therefore, important to investigate
 how the idea of SUSY can be explored in
 future collider experiments
 such as LHC and $e^+e^-$ linear colliders.
 For this purpose, the Higgs sector of
 the SUSY standard models can play
 a unique role. Since the Higgs sector has distinct features,
 its close investigation can give important information on
 these models.

 In the minimal supersymmetric standard model (MSSM) the Higgs sector
 consists of two Higgs doublets, therefore,
 there exist five physical states, i.e. two CP-even Higgs$(h,H)$,
 one CP-odd Higgs$(A)$, and one pair of charged Higgs$(H^\pm)$.
 Since the form of Higgs potential is very restricted
 in the MSSM compared to general two-Higgs models,
 it is possible to derive specific predictions for this Higgs sector.
 For example, the upper bound on
 the lightest CP-even neutral Higgs mass
 is given as about 130GeV\cite{OYY}.
 As for the discovery of the Higgs bosons at
 the future linear collider,
 it is shown that at least one of the CP-even Higgs bosons is
 detectable at an $e^+e^-$ linear collider with
 $\sqrt{s}=300\sim500$GeV \cite{JLC,Janot}.

 Since the discovery of at least one Higgs boson is guaranteed,
 here we would like to address the question of to what extent
 the values of the parameters in the MSSM Higgs sector will be
 determined from the detailed study of the Higgs properties.
 This is especially important in the case
 when only the lightest CP-even Higgs is discovered at a
 future $e^+e^-$ linear collider with $\sqrt{s}=300\sim500$GeV,
 since in such a case the behavior of the Higgs may
 well be quite similar to that of the minimal SM Higgs.
 In order to obtain useful information on the Higgs sector
 it is therefore necessary to measure
 the production cross section and/or
 decay branching ratios precisely and detect possible deviations of
 the Higgs's properties from those of the SM Higgs.
 In the following we consider the determination of
 the parameters in the
 MSSM Higgs sector assuming that only
 the lightest CP-even Higgs is observed
 at a future $e^+e^-$ linear collider with $\sqrt{s}=300\sim500$GeV.
 We show that the determination of the Higgs decay branching ratios in
 the charm and gluonic modes are important to constrain the masses of
 the heavy Higgs sector which may not be observed
 during the earlier stages of the linear collider experiment.
 Therefore precise measurements of branching ratios including
 these modes are useful to set the next beam energy of
 the $e^+e^-$ linear collider at which the heavy Higgs bosons
 can be directly produced.

 Let us begin by listing the parameters in the MSSM Higgs sector
 and the observables available in the experiment
 at the future $e^+e^-$ linear collider.
 Although at the tree level the masses and the mixings of
 the Higgs sector in this model are parametrized by two parameters,
 i.e. CP-odd scalar mass$(m_A)$ and the ratio of
 the vacuum expectation values($\tb\equiv \frac{\langle H_2 \rangle}
 {\langle H_1 \rangle}=\frac{v_2}{v_1}$),
 the radiative correction to the Higgs potential brings
 new parameters into the discussion\cite{OYY}.
 In the calculation of the Higgs effective potential at one loop level
 the most important contribution comes from top and stop loops, and
 therefore the relevant parameters are the
 two stop masses$(m_{\tilde{t_1}}, m_{\tilde{t_2}})$,
 the higgsino mass parameter$(\mu)$ and
 the trilinear soft-breaking mass parameter$(A_t)$.
 For the moment, we assume that no significant effect is induced by
 the left-right mixing of the stop sector.
 Then, the Higgs sector is determined by three parameters,
 which we take to be $m_A$, $\tb$
 and the stop mass scale
 ($m_{susy}\equiv \sqrt{m_{\tilde{t_1}}m_{\tilde{t_2}}}$).
 Note that only this combination of the stop masses
 enters in the Higgs mass formulas through the radiative
 correction as long as the left-right mixing in the stop sector
 is neglected.

 As for the observables, we assume that
 the lightest CP-even Higgs is produced
 through the Higgs bremsstrahlung process, $e^+e^-\to Zh$,
 and that the Higgs decay modes to the SUSY particles are not dominant.
 Then the main decay mode of the lightest Higgs is $h\to b\bar{b}$.
 With reasonable luminosity of
 the $e^+e^-$ linear collider($\sim50$fb$^{-1}$/year)
 we should be able to determine the mass of the Higgs to within
 a few percent\cite{JLC,Janot}.
 We can also expect to measure
 production cross section multiplied by the $h\to b\bar{b}$
 branching ratio, $\sigma(e^+e^-\to Zh)\cdot Br(h\to~b\bar{b})$,
 to within a few percent \cite{JLC,Janot}.\footnote{We can
           also measure the production cross section
           $\sigma(e^+e^-\to Zh)$ by recoil mass distribution
           independently of the Higgs decay modes. This may
           give additional information, but the following discussion
           does not depend on the availability of this quantity.}
 Other measurable quantities are
 branching ratios\cite{Hildreth,Nakamura}.
 The lightest CP-even Higgs has sizable decay
 branching ratios in the modes
 $h\to b\bar{b}$, $\tau\bar{\tau}$,  $c\bar{c}$
 and $gg$\cite{Barger,Yamada,Moretti}.\footnote{
           As stressed in Ref.\cite{Hildreth}, the $h\to WW^{(\ast)}$
           mode is important to distinguish
           the MSSM Higgs from SM Higgs
           for $m_h\gsim 120$GeV.
           Since the branching ratio depends crucially on
           the Higgs mass,
           we will not consider this mode here.}
 Since the $h b\bar{b}$ and $h \tau\bar{\tau}$ couplings originate
 from the Yukawa couplings with the same Higgs doublet,
 the ratio $Br(h\to~\tau\bar{\tau})/Br(h\to~b\bar{b})$ is
 the same as in the SM, and therefore no information on the parameters,
 $m_A,\tb$ and $m_{susy}$ is obtained.\footnote{
           This ratio is important to determine the bottom mass
           as we discuss later.}
 On the other hand, the ratio $Br(h\to c\bar{c})/Br(h\to~b\bar{b})$
 depends on
 these parameters since the charm and bottom couplings to the Higgs
 boson come from Yukawa couplings with different Higgs doublets.
 The dependence of $Br(h\to gg)/Br(h\to~b\bar{b})$ on
 the Higgs parameters is the same as that of
 $Br(h\to c\bar{c})/Br(h\to~b\bar{b})$ in a good
 approximation since the dominant contribution to
 $h\to gg$ is almost always induced by the top quark loop.
 In the following we will consider the quantity
\begin {equation}
     R_{br}\equiv
     \frac{(Br(h\to c\bar{c})+Br(h\to gg))}{Br(h\to b\bar{b})}.
\end {equation}
 It turns out to be possible to determine
 the sum of the charm and gluonic branching ratios to
 a reasonable precision($\pm20\sim25$\%) by the future experiment
 although it is very difficult to measure two branching ratios
 separately with enough precision\cite{Nakamura}.

 Let us now discuss how these three parameters,
 $m_A,\tb$ and $m_{susy}$, will be determined from
 the above observables.
 Since we assume that one CP-even Higgs is observed at
 the $e^+e^-$ linear collider,
 the mass of the Higgs is supposed to be known precisely.
 We then can solve for one of the three parameters in terms of
 the other two using the formula for the lightest CP-even Higgs mass.
 We here solve $\tb$ as a function of $m_A$ and $m_{susy}$
 and try to determine these two parameters from
 the measurements of the production cross section
 and the branching ratios.\footnote{In general,
               there could be two solutions for $\tb$.
               Such a multiple solution, however, occurs only when
               $m_h\lsim 80$GeV and $m_A\lsim 150$GeV.}
 The formulas for the partial decay width of
 the MSSM Higgs is found in
 Ref.\cite{Barger}. QCD corrections are important for the
 $h\to q\bar{q}$ mode\cite{QCDcorr,QCDcrrMeff}
 as well as the $h\to gg$ mode\cite{IKO,SDGZ}.
 For the $h\to q\bar{q}$ partial width we use
 the formula given in Ref.\cite{QCDcrrMeff} where ${\it O}(\alpha_s^2)$
 and $\frac{m_q^2}{m_h^2}$ corrections are taken into account.
 As for the $h\to gg$ mode we use
 the next to leading QCD formula for top loop diagrams\cite{IKO},
 \begin{eqnarray}
    \Gamma(h\to gg) = \Gamma_{LO}[\alpha_s(m_h)]
                   \left( 1+\left( \frac{95}{4}-\frac{7}{6}n_F
                                 \right)\frac{\alpha_s(m_h)}{\pi}
                  \right),
 \end{eqnarray}
 where the leading order formula $\Gamma_{LO}$ is
 found in Ref.\cite{HHG}
 and $n_F=5$. QCD correction in this formula corresponds to
 the case where Higgs mass is far below the $t\bar{t}$ threshold.
 For the Higgs mass region considered here this is
 a good approximation.
 For the other quark's loop we use the leading order result.
 Although the $b$ quark loop can have a sizable contribution for
 large $\tb$, the branching ratio of $h\to gg$ is suppressed
 in such a case. In the parameter region considered in this paper
 contribution from the stop loop is very small and therefore
 is neglected.
 Figure 1 (a) shows the contour plot of $R_{br}$ for $m_h=120$ GeV.
 Here and in the following discussion we take
 the top mass ($m_t$) as 170 GeV,
 the $\overline{\mbox{\rm MS}}$ running quark masses for
 charm and bottom as $\bar m_c(m_c)=1.2$ GeV, $\bar m_b(m_b)=4.2$ GeV
 and the strong coupling constant as $\alpha_s(m_Z)=0.12$.
 Ambiguities associated with these inputs are discussed later.
 A striking feature of this plot is that $R_{br}$ is almost
 independent of  $m_{susy}$.
 This property is useful for constraining the value of $m_A$.
 In figure 1 (b) the contour plot of $\sigma(e^+e^-\to Zh)\cdot
 Br(h\to~b\bar{b})$ are shown for $\sqrt{s}=300$GeV and $m_h=120$ GeV.
 Contrary to the case of $R_{br}$
 the constraint obtained from
 $\sigma(e^+e^-\to Zh)\cdot Br(h\to~b\bar{b})$
 depends on both $m_A$ and $m_{susy}$.
 It may be possible to detect the deviation from
 the SM Higgs by measuring
 this quantity to within a few percent\cite{JLC,Janot,Nakamura},
 however, it is difficult to obtain the constraint on
 $m_A$ independently of $m_{susy}$.

 We can explain the independence of $R_{br}$
 from $m_{susy}$ in the following way.
 The CP-even Higgs mass matrix is given by
 \begin {equation}
 M^2_{ns}=\left( \begin{array}{cc}
         m_A^2\sin^2\beta+m_Z^2\cos^2\beta
         & -(m_A^2+m_Z^2)\cos\beta\sin\beta\\
          -(m_A^2+m_Z^2)\cos\beta\sin\beta
         & m_A^2\cos^2\beta+m_Z^2\sin^2\beta+\frac{\delta}{\sin^2\beta}
       \end{array} \right)\label{eqn:matrx},
 \end {equation}
 where
 \begin {equation}
 \delta=\frac{3}{4\pi^2}\frac{m_t^4}{v^2}
         \ln\left(\frac{m_{susy}^2}{m_t^2}\right)
 \end {equation}
 represents the top-stop loop effect in the calculation of
 the Higgs effective potential. From this matrix
 the Higgs mixing angle, $\alpha$, is given by
 \begin {equation}
   \tan\alpha=
   \frac{(m_Z^2+m_A^2)\sbt\cb}{m_h^2-(m_Z^2\cbsq+m_A^2\sbsq)}.
  \label{eqn:tana}
 \end {equation}
 Since the dependences of the branching ratios
 on the angles $\alpha$ and $\beta$ are given by
 \begin {equation}
   Br(h\to b\bar{b})\propto\frac{\sin^2\alpha}{\cbsq}
 ,\ \ \   Br(h\to c\bar{c})\propto\frac{\cos^2\alpha}{\sbsq},
 \end {equation}
 the ratio of $Br(h\to b\bar{b})$ and $Br(h\to c\bar{c})$
 is proportional to
 \begin {equation}
   \frac{Br(h\to c\bar{c})}{Br(h\to b\bar{b})}
   \propto\left(\frac{1}{\tb\tan\alpha}\right)^2.
 \end {equation}
 By using Eq.\ref{eqn:tana},
 $\frac{1}{\tan\beta\tan\alpha}$ is rewritten as
 \begin {equation}
     \frac{1}{\tb\tan\alpha}=\frac{m_h^2-m_A^2}{m_Z^2+m_A^2}\left\{
             1+\frac{m_h^2-m_Z^2}{m_h^2-m_A^2}\frac{1}{\tbsq}\right\},
  \label{eqn:tatb}
 \end {equation}
 where the $m_{susy}$ dependence is implicit in $\tb$.
 Since the second term in the parenthesis in
 Eq.\ref{eqn:tatb} is negligible for $m_A^2 \gg m_h^2\sim m_Z^2$,
 Eq.\ref{eqn:tatb} is approximately given by
 \begin {equation}
     \frac{1}{\tb\tan\alpha}\approx\frac{m_h^2-m_A^2}{m_Z^2+m_A^2}.
 \label{eqn:tatbappr}
 \end {equation}
 This explains the independence from $m_{susy}$ for the ratio of
 the $h\to b\bar{b}$ and $h\to c\bar{c}$ branching ratios.
 Under the present assumption that the only lightest CP-even
 Higgs boson is discovered at an $e^+e^-$ linear collider
 with $\sqrt{s}=300\sim500$GeV, the above mass relations among
 $m_A,  m_h$, and $m_Z$ are naturally satisfied.
 For the $h\to gg$ mode, the situation is similar to
 $Br(h\to c\bar{c})/Br(h\to b\bar{b})$ because
 the dominant contribution to the $h\to gg$ mode
 almost always comes from
 the top-loop diagram and $Br(h\to gg)$ has
 the same $\alpha$ and $\beta$ dependence as $Br(h\to c\bar{c})$.
 This is the reason why $R_{br}$
 is almost independent of $m_{susy}$.

 Let us next consider how well $m_A$ will be constrained from
 the measurement of
 $R_{br}$. In figure 2 (a) we show the $R_{br}$
 as a function of $m_A$ for $m_{susy}=0.75, 1, 5, 10$ TeV and
 $m_h=120$ GeV. Figure 2 (b) corresponds to the case for
 $m_{susy}=0.5, 1, 2, 10$ TeV and $m_h=100$ GeV.
 In these figures some of the lines terminate because
 we cannot obtain a solution beyond the end point for
 the assumed Higgs mass.
 Although $R_{br}$ becomes
 the SM value in the limit $m_A\to \infty$,
 we can see that the ratio is about 20\% smaller than
 the SM value even at $m_A=400$GeV.
 Note that the direct search for the CP-odd Higgs can exclude
 only the mass region approximately half of the $\sqrt{s}$
 (i.e. $m_A\lsim 250$GeV for $\sqrt{s}=500$GeV)
 since the associated production of $A$ and
 the heavy CP-even Higgs $H$
 is the only practical production mechanism in this region.
 Therefore $R_{br}$ can be a good probe into the heavy Higgs bosons
 in the mass region larger than $\frac{\sqrt{s}}{2}$.
 For example, if the experimental result is given by
 $R_{br}=0.10\pm0.02$,
 we will be able to constrain the value of $m_A$ to
 260~GeV~$\lsim~m_A~\lsim~400$~GeV.
 On the other hand, in the case that the branching ratio is consistent
 with the SM value, the lower bound on $m_A$ may be obtained.

 In the above discussion we have assumed that the left-right mixing in
 the stop sector is negligible. This is especially used in deriving
 Eq.\ref{eqn:tatb}. To see how the above results depend on
 the mixing effects
 we have calculated
 $R_{br}$ using a one loop potential including the trilinear soft
 breaking term for the stop ($A_t$ term) as well as the supersymmetric
 higgsino mass term ($\mu$ term) as in Ref.\cite{Ellis}.
 Examples are shown in figure 3. We can see that for reasonable
 values of  $A_t$ and $\mu$,
 $R_{br}$ does not strongly depend on the parameters
 in the stop sector.\footnote{As pointed out in Ref.\cite{Kane},
         it is possible to suppress the $hb\bar{b}$ coupling using
         the left-right mixing effect.
         But this only occurs for special choices of parameters
         for which the effect of $A_t$ and $\mu$
         is large enough to cancel
         the off-diagonal term in Eq.\ref{eqn:matrx}.}
 Sbottom loops can also affect the Higgs mass formulas for
 large values of $\tb$.
 As long as the left-right mixing in the sbottom sector is neglected,
 this effect is negligible for $R_{br}$.
 In such a case a term
 $\frac{\delta_b}{\cos^2\beta}$ is added in the (1,1)
 component of the CP-even Higgs matrix in Eq.\ref{eqn:matrx},
 where
 $\delta_b=
 \frac{3}{4\pi^2}\frac{m_b^4}{v^2}\ln{\frac{m_{sbottom}^2}{m_b^2}}$,
 and Eq.\ref{eqn:tatb} is modified as
 \begin{eqnarray}
     \frac{1}{\tb\tan\alpha}=\frac{m_h^2-m_A^2}{m_Z^2+m_A^2}
            \left\{
               1+\frac{m_h^2-m_Z^2}{m_h^2-m_A^2}\frac{1}{\tbsq}
       -\frac{\delta_b}{m_h^2-m_A^2}
        \frac{(1+\tan^2\beta)^2}{\tan^2\beta}\right\}
  \label{eqn:tatb2}
 \end{eqnarray}
 Even for $\tb\simeq \frac{m_t}{m_b}$ the last term in the parentheses
 is suppressed by $\frac{3}{4\pi^2}\frac{m_b^2}{m_h^2-m_A^2}$ and
 $R_{br}$ is almost independent of the SUSY breaking scale.

So far we have neglected uncertainties in the quark masses and
the strong
coupling constant in the calculation of the branching ratios.
The precise determination of the $\overline{\mbox{\rm MS}}$
running quark mass ratio $\bar m_c^2(m_h)/\bar m_b^2(m_h)$ would be
especially important in the calculation of $R_{br}$. For this purpose
we need input parameters for charm and bottom quark masses at some
renormalization scale. We may use results from several non-perturbative
methods such as lattice QCD, QCD sum rule \cite{NA} and the heavy quark
effective theory(HQET). As an illustration
we estimate the uncertainties
in $\bar m_c^2(m_h)/\bar m_b^2(m_h)$ using the results of lattice QCD
and HQET. In Ref.\cite{CGMS} the $\overline{\mbox{\rm MS}}$ bottom
quark mass at the bottom scale is given as
$\bar m_b(m_b)=(4.17\pm 0.05\pm 0.03)$GeV
by lattice QCD. On the other hand HQET gives the difference of the
charm and bottom quark mass,
which should be understood as {\em pole masses},
as $\Delta M_{bc}=M_b-M_c=(3.40\pm 0.03\pm 0.03)$GeV \cite{N}.
Note that although the pole mass itself has no physical meaning in QCD
the difference does have at least in the leading order of the $1/m$
expansion\cite{S}. From these two inputs and
the two-loop relation between
the perturbative pole mass and the $\overline{\mbox{\rm MS}}$
mass\cite{T,GL}%
\footnote{Although a three-loop result has already appeared in
the literature\cite{GBGS}, we use the two-loop results throughout
this illustration for the consistency with the remaining part of
the calculation.},
 $M_Q={\overline m}_Q(M_Q)
    \left(1+\frac{4}{3}\frac{\alpha_s(M_Q)}{\pi}\right),$
we can calculate the $\overline{\mbox{\rm MS}}$
running quark masses.
Using $\alpha_s(m_Z)=0.117
\pm 0.006$\cite{B} in addition to the above two inputs we obtain
\begin{equation}
\frac{\bar m_c^2(m_h)}{\bar m_b^2(m_h)}=
                        0.026\pm 0.004
                             \pm 0.004
                             \pm 0.003,\\
\label{R}
\end{equation}
where the first (second, third) error comes from $\alpha_s(m_Z)~
(m_b,\Delta M_{bc})$.
 Uncertainty in $Br(h\to gg)/Br(h\to b\bar{b})$ comes from
top and bottom masses and $\alpha_s$ as well as
 the next-to-next-to-leading order QCD corrections
 in the gluonic partial width.
 If the top and bottom quark masses are known precisely,
 the strong coupling constant gives the largest uncertainty.
 Varing the strong coupling constant as
 $\alpha_s(m_Z)=0.117\pm0.006$, the gluonic partial width changes
 by $\pm12\%$.
              \footnote{Recently, it is  pointed out that
                the uncertainty in the strong coupling constant
                causes  errors in the theoretical
                predictions of the branching ratios of the modes
                $h\rightarrow c\bar c,gg$ as large as 20 \%
                \cite{DSZ}. Our estimation is consistent with
                their result. As is stated in the text, however,
                this uncertainty in the strong coupling constant
                could reduce
                much in the linear collider era.}

 Although the present uncertainty is relatively large, we can
 expect theoretical and experimental improvements in future.
 For example the $\alpha_s$ measurement at the $t\bar{t}$ threshold
 is expected to reduce the error in $\alpha_s(m_Z)$ by
 factor of 3 at the $e^+ e^-$ linear collider experiment\cite{FMS}.
 Also direct measurement of the $\overline{\mbox{\rm MS}}$
 running bottom
 quark mass at the scale of $m_h$ may be possible from
 the measurement
 of $Br(h\to \tau\bar{\tau})/Br(h\to b\bar{b})$\cite{JLC}.

 To summarize, we have examined
 whether the parameters in the MSSM Higgs
 sector can be determined by detailed study of the Higgs properties.
 We pointed out that the ratio of
 $Br(h\to c\bar{c})+Br(h\to gg)$ and $Br(h\to b\bar{b})$ is
 sensitive to the heavy Higgs mass scale but its dependence on
 the stop mass scale is very weak. Therefore, if this ratio is
 measured with enough precision at
 the future $e^+ e^-$ linear collider,
 we may be able to constrain the heavy Higgs mass scale even if
 the heavy Higgs bosons cannot be produced directly.\\

 The authors would like to thank A. Djouadi, K. Hagiwara, K. Kawagoe,
 I. Nakamura, M. Peskin and P.M. Zerwas for useful discussions.
 They also wish to thank B. Bullock and K. Hikasa for
 reading the manuscript
 and useful comments.
 This work is supported in part by the Grant-in-aid for Scientific
 Research from the Ministry of Education, Science and Culture of
 Japan.

\newpage
\baselineskip24pt
\noindent{\large {\bf Figure Captions}}\\
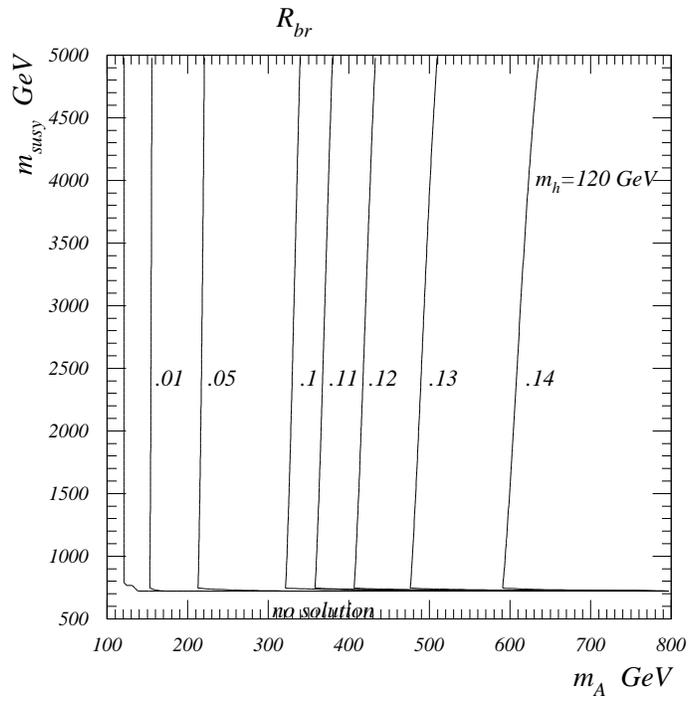
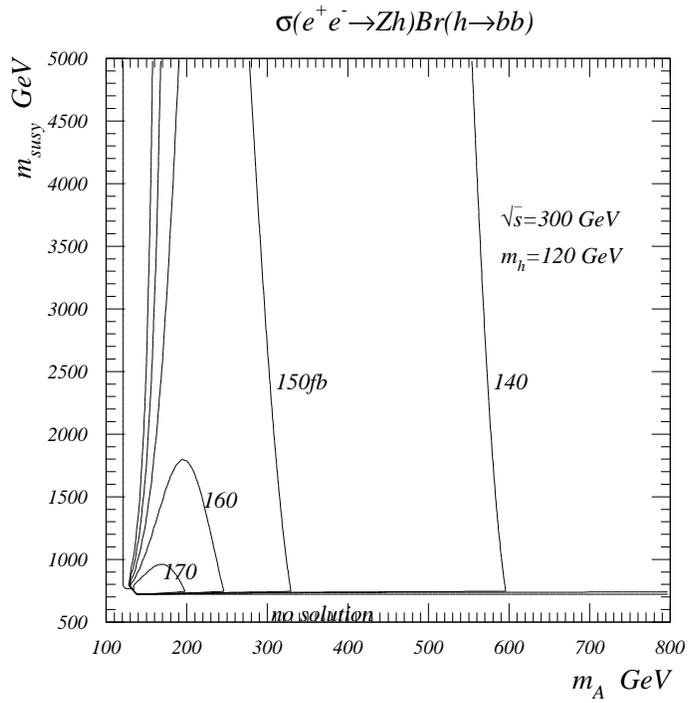
Fig.1: (a) $R_{br}\equiv
\frac{(Br(h\to c\bar{c})+Br(h\to gg))}{Br(h\to b\bar{b})}$
in the parameter space of the CP-odd Higgs mass ($m_A$) and
the stop mass
scale ($m_{susy}$) for the lightest CP-even Higgs mass $m_h=120$ GeV.
We have used $m_t= 170$ GeV, $\bar m_c(m_c)=1.2$ GeV,
$\bar m_b(m_b)=4.2$ GeV and $\alpha_s(m_Z)=0.12$.
(b) $\sigma(e^+e^-\to Zh)\cdot
 Br(h\to~b\bar{b})$ for $\sqrt{s}=300$GeV.
 Other parameters are the same as in (a).\\

\noindent
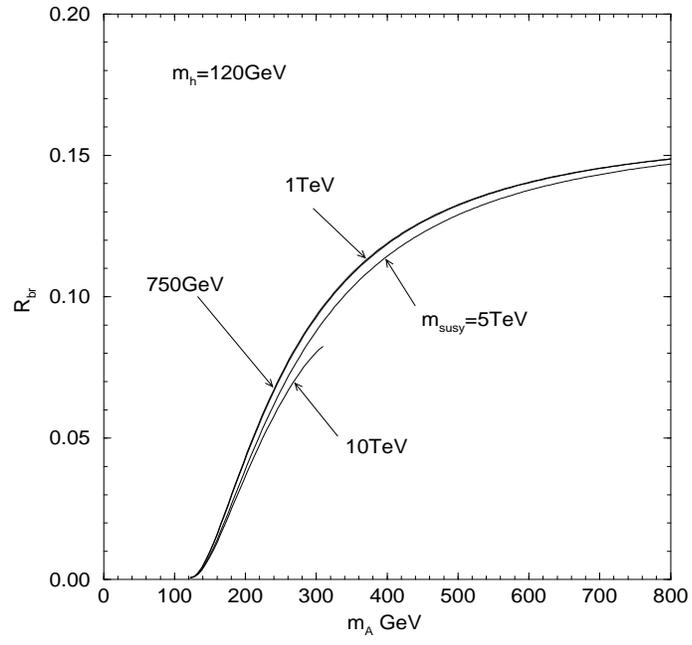
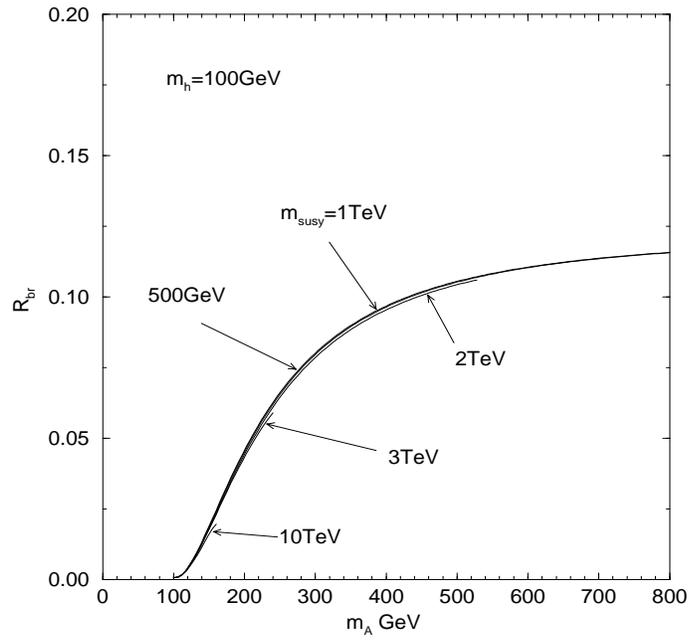
Fig.2: $R_{br}$ as a function of $m_A$ for several values of $m_{susy}$
for $m_h=120$ GeV (a) and for $m_h=100$ GeV (b). Other input parameters
are the same as those for figure 1 (a).\\

\noindent
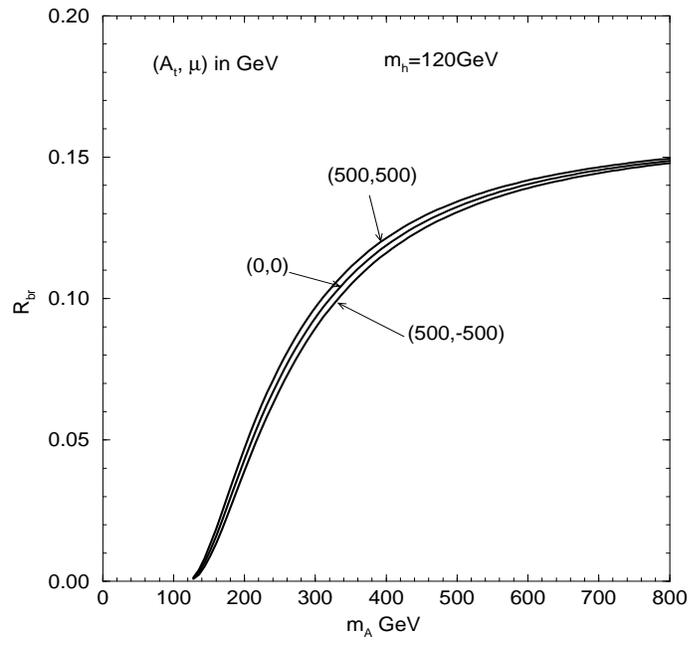
Fig.3:  $R_{br}$ as a function of $m_A$ including the left-right mixing
effects of two stops. We take $m_h=120$ GeV and two stop masses
as $m_{\tilde{t_1}}=1$ TeV and $ m_{\tilde{t_2}}=700$ GeV. The values
shown in the parentheses represent ($A_t,\mu$) in GeV.

\end{document}